\documentclass[prd,nofootinbib,twocolumn,showpacs]{revtex4}
\usepackage{graphics}
\usepackage{bm}

\usepackage{amsmath}    
\usepackage{amssymb}    
\usepackage{graphicx}   
\usepackage{epsf}
\usepackage{epsfig}
\begin{document}

\title {Spontaneous Fluxon Production in Annular Josephson Tunnel Junctions in the Presence of a Magnetic Field}
\thanks{Submitted to Phys. Rev. B.}

\author{R.\ Monaco}
\affiliation{Istituto di Cibernetica del C.N.R., 80078, Pozzuoli,
Italy and Unita' INFM-Dipartimento di Fisica, Universita' di
Salerno, 84081 Baronissi, Italy.}\email
{roberto@sa.infn.it}
\author{M.\ Aaroe and J.\ Mygind}
\affiliation{Department of Physics, B309, Technical University of
Denmark, DK-2800 Lyngby, Denmark.} \email{aaroe@fysik.dtu.dk}
\author{R.\ J.\ Rivers}
\affiliation{Blackett Laboratory, Imperial College London, London
SW7 2AZ, U.K. }\email{r.rivers@imperial.ac.uk}
\author{V.\ P.\ Koshelets}
\affiliation{Institute of Radio Engineering and Electronics,
Russian Academy of Science, Mokhovaya 11, Bldg 7, 125009, Moscow,
Russia.}\email{valery@hitech.cplire.ru}
\date{\today}

\begin{abstract}
We report on the spontaneous production of fluxons in annular
Josephson tunnel junctions during a thermal quench in the presence
of a symmetry-breaking magnetic field. The dependence on field
intensity $B$ of the probability $\bar{f_1}$ to trap a single
defect during the N-S phase transition  depends drastically on the
sample circumferences. We show that this can be understood in the
framework of the same picture of spontaneous defect formation that
leads to the experimentally well attested scaling behaviour of
$\bar{f_1}$ with quench rate in the absence of an external field.
\end{abstract}


\maketitle

\section{Spontaneous fluxon production in an external field}

Some time ago it was proposed by Kibble \cite{kibble1} and Zurek
\cite{zurek1,zurek2} that the domain structure after a continuous
phase transition is determined by its causal horizons. Since then
there has been a sequence of experiments in condensed matter systems
to attempt to confirm this by counting the topological defects
produced at a quench, whose number can be correlated to phase
boundaries.  Such confirmation lies in the predicted scaling
behaviour of the average defect separation ${\bar\xi}$ at the time
of their production as a function of the quench time (inverse quench
rate at the critical temperature $T_{c}$) $\tau_Q$. Supposing that
the 'equilibrium' correlation length $\xi _{eq}(T)$ of the
order-parameter field, and its relaxation time $\tau (T) $, diverge
at $T=T_{c}$ as:
\begin{equation}
\xi _{eq}(T)=\xi _{0}\bigg|1-\frac{T}{T _{c}}\bigg|^{-\nu },\,\,\tau
(T)=\tau _{0}\bigg|1-\frac{T}{T _{c}}\bigg|^{-\gamma }, \nonumber
\end{equation}
\noindent simple causal arguments predict behaviour of the
form\cite{kibble1,zurek1,zurek2}:
 \begin{equation}
 {\bar\xi}\propto \xi_0(\tau_Q/\tau_0)^{\sigma},\label{xibar}
 \end{equation}
 where the scaling exponent $\sigma > 0$ belongs to universality
 classes determined by the (adiabatic) critical exponents $\nu$ and $\gamma$,
 and $\nu$, $\gamma$, $\xi_0$ and $\tau_0$ depend on the detailed
 microscopic behaviour of the system. The coefficient of proportionality in (\ref{xibar}) is an
efficiency factor which varies with the system, from a few percent
for high temperature superconductors \cite{carmi,technion2} to
full efficiency for superfluid $^3He$ \cite{grenoble,helsinki}.

In the last several years we have performed a set of experiments
\cite{Monaco1,Monaco2,Monaco3,Monaco4} on planar annular Josephson
Tunnel Junctions (JTJs) to test the scaling law (\ref{xibar}).
Specifically, a planar Josephson junction comprises two
superconducting films separated by an
insulating oxide layer. We assume continuity in the density of
Cooper pairs across the oxide, but allow for a discontinuity $\phi$
in the phase of the effective order parameter field. Once the
transition is completed the lossless Josephson current density is
 $J=J_c\sin\phi$, for critical current density $J_c$.
For such JTJs the defects are fluxons (or antifluxons)
corresponding to a change $\Delta\phi = \pm 2n\pi$ in $\phi$ along
the oxide layer \cite{barone}. The integer $\pm n$ is the
so-called winding number. The Swihart velocity provides the
requisite causal horizons \cite{KMR}.

 Most simply, for small annuli of circumference $C\ll{\bar\xi}$, the trapping probability
 $f_1$ for finding a fluxon (or $f_{-1}$ of finding an antifluxon)
 is taken to be:
 \begin{equation}
 f_1 = f_{-1} = C/{\bar\xi} \propto
 (C/\xi_0)(\tau_Q/\tau_0)^{-\sigma}.\label{flin}
 \end{equation}

In our experiments we measure ${\bar f}_1 = f_1 + f_{-1}$, the
likelihood of seeing one fluxon or antifluxon. We have carried out
statistical measurements to determine the dependence of
$\bar{f_1}$ on $\tau_Q$ for a large number of high quality AJTJs
with circumferences varying from $0.5 \,mm$ to $3.14 \,mm$. All
samples had equal critical current density $J_{c}(T=0)\simeq
60A/cm^2$ (corresponding to a Josephson penetration depth
$\lambda_{J}(T=0)\simeq 50\mu m$), yielding the same $\xi_0$ and
$\tau_0$. The quenching time $\tau_Q$ could be changed over
several orders of magnitudes from tens of milliseconds to tens of
seconds, using the methods described in \cite{Monaco3,Monaco4}.

  In Fig. 1 we show  ${\bar f}_1(\tau_Q)$ for annuli of radius $0.5 mm$ (lower plot)
and $1.5 mm$ (upper plot). The results for the larger annulus have
not been shown before.  There is no doubt that, for small
$C/{\bar\xi}$, the probability ${\bar f}_1$ of finding a single
fluxon shows scaling behaviour of the type (\ref{flin}), with
$\sigma = 0.5$ to high accuracy, some of whose details were given in
\cite{Monaco3,Monaco4}. This value of $\sigma$ is as we would expect
for realistic junctions for which the fabrication leads to a
proximity effect \cite{Rowell&Smith,golubov} and for which the
Swihart velocity does not show complete slowing down.  The
efficiency factor is approximately unity for the smaller annulus.
More details are given in \cite{Monaco4}.

 \begin{figure}
 \hspace{2cm}\scalebox{0.8}
 {\includegraphics{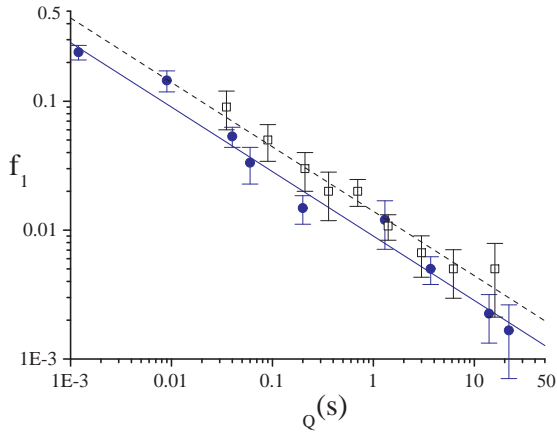}}
\caption{Log-log plot of the frequency ${\bar f}_{1}$ of trapping
single fluxons versus the quenching time $\tau _{Q}$ for an AJTJ of
circumference $0.5 mm$ (closed circles) and for an AJTJ with
circumference $1.5 mm$. The best fits through the data (solid and
dashed lines) show that scaling with $\sigma = 0.5$ is totally
robust. Both samples had equal critical current density $J_{c}$.}
\label{Fig.1}
\end{figure}

\noindent What concerns us in this paper is how fluxons form in the
presence of an externally applied magnetic field $B$ that explicitly
breaks the symmetry of the theory, whereby $f_1\neq
 f_{-1}$. This is a crucial ingredient in the analysis of {\it unbiased} fluxon production
in JTJs because, despite our best efforts, we cannot preclude the
possibility of stray magnetic fields in the experimental equipment
(e.g. a magnetised screw or an incomplete shielding of the earth's
magnetic field). In fact, in presenting the data in Fig.1 we have
taken the effects of static stray fields empirically into account,
by applying an external field until such stray fields are
neutralised. It is only then that we obtain (\ref{flin}). In all
cases the external magnetic field $B$ was applied perpendicular to
the junction plane. This choice of field orientation is mainly due
to the fact that a transverse field (due to demagnetization
effects) is more effective, by almost two orders of magnitude,
than an in-plane field in modulating the junction critical current
$I_c$ \cite{Monaco5} and trapping frequency. Furthermore, under
particular conditions \cite{Monaco6} a transverse magnetic field
allows to discriminate between fluxons and antifluxons.

As a result, we have built up a substantial collection of data
showing the dependence of ${\bar f}_1$ on {\it both} $\tau$ and
$B$, that we shall discuss in the remaining  sections. In
particular, we shall concentrate on two representative datasets
that refer to high quality $Nb/Al_{ox}/Nb-Nb$ AJTJs quenched at
the same quench rate ($\tau_Q=5s$), but having different
circumferences, i.e., $C=0.5mm$ and $C=2.0mm$, shown in Fig.2 and
Fig.3, respectively. Details of the samples' electrical and
geometrical parameters and of the experimental setup can be found
in \cite{Monaco4}.

\begin{figure}
\hspace{2cm}\scalebox{0.8}
 {\includegraphics{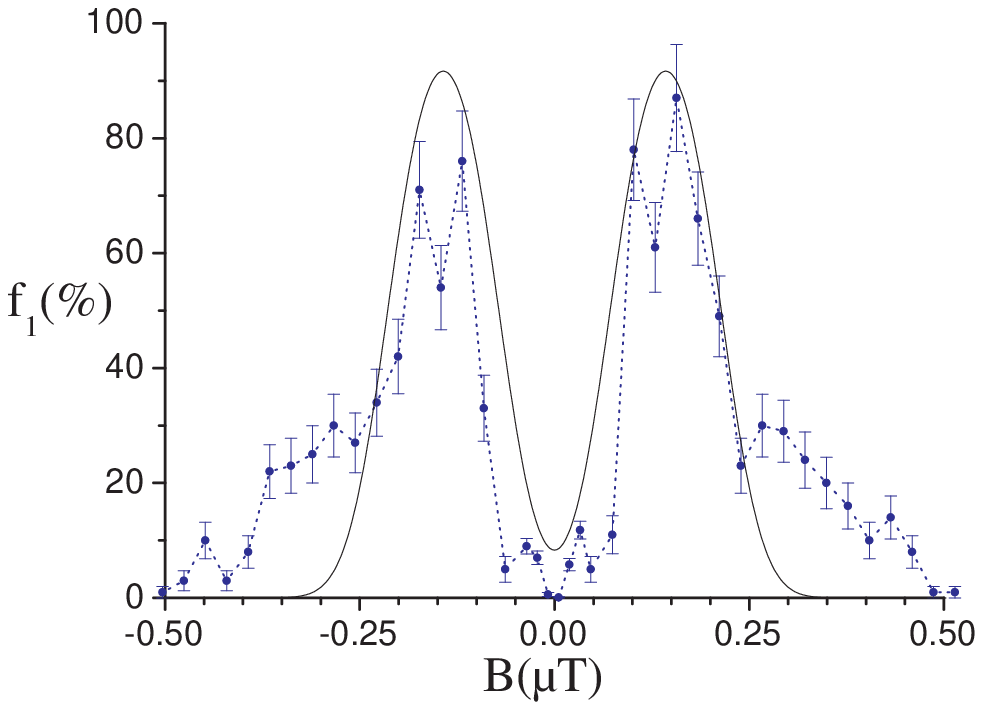}}
\caption{Dependence of the single fluxon trapping frequency
$\bar{f_1}$ for quench time $\tau_Q=5s$ for an AJTJ of circumference
$C=0.5mm$ in the presence of a magnetic field $B$ perpendicular to
the barrier plane. The solid line corresponds to the case $N=1$ in
${\bar f}_1(N, {\bar n})$ of Fig.5, as derived from Eq.(\ref{fpq}).}
\label{Fig.2}
\end{figure}

\begin{figure}
\hspace{2cm}\scalebox{0.8}
 {\includegraphics{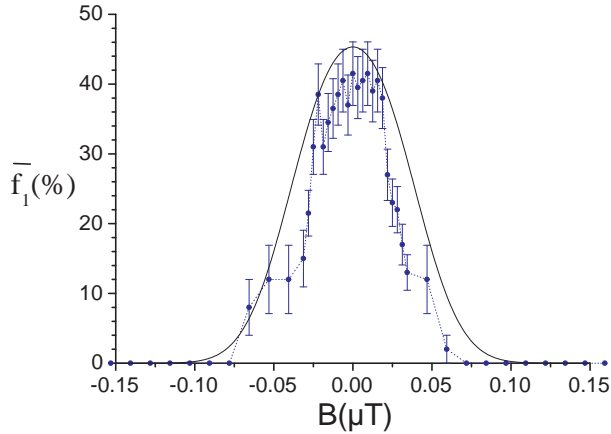}}
\caption{Dependence of $\bar{f_1}$ for quench time $\tau_Q=5s$ for
an AJTJ of circumference $C=2.0mm$ in the presence of a magnetic
field $B$ perpendicular to the barrier plane. The solid line
corresponds to the case $N=16$ in ${\bar f}_1(N, {\bar n})$ of
Fig.5, as derived from Eq.(\ref{fpq}).} \label{Fig.3}
\end{figure}

We observe that the increase in circumference has a dramatic
effect on the single trapping frequency ${\bar f}_1$. For the
$0.5mm$ long AJTJ we find a central minimum with two side-peaks.
Such a double-peaked dataset has been used for any single data
point in Fig.1, with ${\bar f}_1$ being read off from the central
minimum, typically displaced slightly from $B=0$ because of the
aforementioned stray fields in the equipment (typically several
tens of $nT$).

On the other hand, for the $2.0mm$ long sample (not represented in
Fig.1) we only have a central peak. Superficially this is strange,
since it shows that the probability of seeing a single fluxon {\it
decreases} as the external field increases, but we shall
understand this as a consequence of an increased ability to create
more than one fluxon. We note that the values of the magnetic
field required to change ${\bar f}_1$ significantly are very small
when compared to the field values needed to modulate the Josephson
current $I_c$ of the samples, whose first minimum occurs at field
values of several $\mu T$. Further, the larger the ring size the
larger is the effect of a given magnetic field.

In the remainder of this paper we shall show how the results of Figs. 2 and 3
can be understood, and qualitatively predicted, in a framework that implies the scaling behavior of Fig.1.
As such, these results are mutually supportive but require an alternative formulation of the KZ scenario,
which we now provide.

\section{Causality vs. instability}

At first sight the results for fluxon production in an external
field have little or nothing to do with the Kibble-Zurek scenario.
However, the original scenario of bounding domains by causal
horizons is just one way of saying that, qualitatively, systems
change as fast as is possible. From a different viewpoint we know
that continuous transitions (like the one here) proceed by the
exponential growth of the amplitudes of unstable long-wavelength
modes of the system. This growth is strongly suppressed by
self-interaction once the system is close to the ground states of
its symmetry-broken phase.

Exponential growth is as fast as it gets, corresponding to
linearising the equations of motion. Thus, provided there is
enough time for such rapid growth before back-reaction
(self-interaction) stops it, we can understand how systems can
change as fast as is possible without invoking causal horizons
directly. There is a corollary to this. The linear behaviour of
the order parameter fields, while the transition is taking place,
follows from their behaving as Gaussian random variables. In fact,
for an idealised situation in which amplitude growth is stopped by
implementing a rapid back-reaction that chokes off growth
instantaneously, it can be shown \cite{karra,ray,bowick} that the
assumption of Gaussianity leads to  scaling behaviour like that of
(\ref{xibar}), with the same values of $\sigma$ as would be
obtained from causal bounds. What may seem surprising is that,
even when back-reaction is included more realistically, numerical
simulations \cite{lag,moro,ray2} show that the behaviour is still
essentially the same. The $\sigma$ exponents of causal reasoning
are recovered. There are, however, two major differences between
fastest amplitude growth and causal bounds. The growth starts {\it
after} the transition has begun, whereas causal bounds can be
imposed both {\it before} or {\it after} the transition has begun,
usually to the same effect \cite{zurek1,zurek2}. Numerical
simulations show \cite{moro,ray2}, without a doubt, that it is the
behaviour of the system {\it after} the transition has begun that
determines the domain structure. This is necessary in our context
since, with no Josephson effect for $T>T_c$, we could not have
invoked causality before the transition. Further, the assumption
of Gaussianity gives us more than causality, in that it
reintroduces the role of the Ginsburg temperature, at which
thermal fluctuations become important,where appropriate. This
provides one natural explanation for the failure to observe
vortices in quenches of $^4He$ \cite{ray} while, by a similar
argument, permitting spontaneous vortex production in $^3He$
\cite{grenoble,helsinki}  and superconductors \cite{technion2}.

However, since equations of motion are, by construction, causal, these viewpoints are
largely complementary once these caveats are taken into account.
When appropriate we will invoke both mechanisms in our subsequent discussion.

The simulations cited above are largely for systems with global
symmetry breaking, simpler than superconductors. In fact, although
local breaking gives a very different domain structure, the idea
of the transition being driven by instabilities survives. Again,
the exponents are those of causal arguments \cite{calzetta} except
that there is a {\it additional} mechanism\cite{rajantie} for the
spontaneous production of flux in which magnetic field just
freezes in by itself, with behaviour very different to that of
(\ref{xibar}). For annular JTJs this further mechanism does not
arise because of the thinness of the oxide layer through which
fluxons protrude and the scaling behaviour of (\ref{xibar}) and
(\ref{flin}) is a clean prediction.

\section{Beyond The Linear Regime ($B=0$)}

Suppose that there is no external symmetry-breaking field. As long
as $f_1$ is significantly smaller than unity we expect the {\it
linear} log-plot in $\tau_Q$, as seen in Fig.1. However, with
$f_1$ bounded by unity the linear behaviour will soon break down
for increasingly fast quenches. Once ${\bar\xi}= O(C)$ the
trapping probability $f_m$ of finding net fluxon number $m$
(fluxons minus antifluxons)  increases for $m>1$, forcing $f_1$ to
decrease. For the remainder of this section we shall extend
(\ref{flin}) to predictions for $f_1$ and higher $f_m$ across the
whole range of $C/{\bar\xi}$.

Since the KZ scenario is appropriate for $B=0$, this is an ideal
testing ground for our two approaches. We shall elaborate on these
in turn and see that, for many purposes, they give almost
indistinguishable results.

\subsection{Independent domains and Gaussian probabilities}

In the spirit of the KZ scenario a simple, but informative first
guess as to how $f_1$ and other $f_m$ behave across the whole
range of $C/{\bar\xi}$ is to divide the annulus into $N\geq 2$
independent (causal) domains in each of which the Josephson phase
$\phi$ is a constant. We assume that there is no correlation
between the values of $\phi$ in adjacent domains but, in
calculating the total phase change $\Delta\phi$ around the
annulus, the geodesic rule is adopted \cite{rudaz}. This means
that, when jumping from one domain to the other, the shortest path
in phase will be taken. The result of a quench is then modelled as
having the system divided up into N domains, each with a randomly
chosen phase. There is nothing in this ansatz peculiar to JTJs,
and it is equally applicable to superconductors. As such it is an
idealisation of a superconducting loop, made out of Josephson
junctions in series, that has been the object of spontaneous flux
generation \cite{carmi}.

Let $G_M(\Delta\phi)$ be the probability that the change in phase
$\phi$ is $\Delta\phi$ after $M$ domain boundaries. If the system is
made of only two domains, then the lack of phase correlation
requires:
\begin{eqnarray}
 G_1(\Delta\phi) &=& \frac{1}{2\pi}\, \,\,\,\,\,for\,\,\, -\pi < \Delta\phi < \pi,\nonumber\\
&=& 0 \,\,for\,\,\, |\Delta\phi |>\pi\nonumber.
\end{eqnarray}
 In fact, the approach of using the KZ
picture to set up discrete domains in which the order parameter
field can take random values is one that has been used repeatedly
for counting defects, at least since its introduction by
Vachaspati and Vilenkin \cite{tanmay} for counting cosmic strings
(vortices) in the early universe. This adopts an earlier use of
causal horizons by Kibble \cite{kibble2}, from which
\cite{kibble1} evolved.

\noindent On increasing $N$, $G_N(\Delta\phi)$ is determined by
$N-1$ self-convolutions of $G_1$:
\begin{equation}
 G_N = \underbrace{G_1 *...*G_1}_{N-1\,\,\, times}.
  \label{GN}
  \nonumber
 \end{equation}
 Applying the geodesic rule for the final step in phase (from
 domain $N$ back to domain $1$, the probability of ending with a
 phase shift of $2\pi m$ (i.e. net fluxon number $m$) is:
 \begin{equation}
 f_m(N) = \int_{-\pi + 2m\pi}^{\pi + 2m\pi}d\Phi\,G_N(\Phi) =
 2\pi G_{N+1}(2m\pi), \label{fmN}
 \end{equation}
 where the last equality comes from the definition of $G_N$.

  To bring this further into correspondence with the KZ scenario, we should identify
 the domain size as comparable to $C/{\bar\xi}$ i.e. $C = aN{\bar\xi}$,
 where $a = O(1)$. The value of $a$ is not unity, since a discrete
 domain structure is only a crude approximation to a continuous
 phase at a continuous transition. Further, the result
 (\ref{xibar}) assumes the causal bound is saturated, and we have
 already commented on systems (e.g. high-$T_c$ superconductors) for which the
 inefficiency of producing flux shows that this is not the case \cite{technion2}.
 To take general values of $C$ and ${\bar\xi}$ into account we
 need to generalise (\ref{fmN}) to non-integer $N$.

$G_N(\Delta\phi)$  already
 shows rapid convergence to the Gaussian distribution that arises
 from the central limit theorem for $N\geq 2$.
 For such $N$
the obvious way to proceed is to adopt this central limit Gaussian
distribution. That is, we assume that the total phase change
$\Delta\phi$ around the annulus can be expressed as the sum of a
random term $\Phi$ and a geodesic-rule correction $\delta\Phi$. If
$\Phi$ has a normal distribution with average $\bar \Phi = 0$ and
variance $\sigma^2(N)$ i.e.:
 \begin{equation}
 G_N(\Phi) = \frac{1}{\sqrt{2 \pi \sigma^2(N)} }
\exp-\bigg({\frac{\Phi^2}{2 \sigma^2(N)}}\bigg),
 \label{Ggauss}
 \end{equation}
then a simple calculation enables us to identify (\ref{Ggauss}) with
the central limit distribution of the self-convolutions of $G_1$, as
described above, provided that:
  \begin{equation}
 \sigma^2 (N) = N\pi^2/3.
 \label{sigmaN}
  \nonumber
 \end{equation}
The probability to trap a net number $m$ of defects will now be:
\begin{equation}
f_{m}(N) =  \int_{-\pi + 2m\pi}^{\pi + 2m\pi}d\Phi\,G_N(\Phi).
\label{fpm}
\end{equation}

Experimentally, in the absence of any external field, the most
important probabilities are for finding one fluxon or one
antifluxon. The frequency of no trapping $f_0(N)$ is:

\begin{equation}
f_0(N) = \int_{-\pi}^{\pi} G_N(\Phi) d\Phi = erf(\frac{\pi}{\sqrt
{2 \sigma^2(N)} }),
 \label{fp0}
 \end{equation}

and

\begin{eqnarray}
f_{\pm 1}(N) &=&  \int_{\pi}^{3\pi}\,\frac{d\Phi}{\sqrt{2 \pi
\sigma^2(N)} } \exp-\bigg({\frac{\Phi^2}{2 \sigma^2(N)}}\bigg).
 \label{fp1}
\end{eqnarray}
\noindent Furthermore, for large $ \sigma^2>>\frac{\pi^2}{2}$, say $ \sigma^2 \geq 20$, the trapping
frequencies asymptotically approach zero as:

\begin{equation}
 f_{\pm m}(N) \approx  \sqrt{\frac{2\pi}{\sigma^2(N)}} \exp
 -\bigg(
\frac{2 \pi^2 m^2}{\sigma^2(N)}\bigg).
\end{equation}

\noindent Finally, in the same limit, the variance of the discrete
variable $n$ is:
 \begin{equation}
 \sigma_n^2(N) = \Sigma_{n=-\infty}^\infty
n^2 f_n(N) = <n^2>= \sigma^2(N)/4\pi^2.
 \end{equation}

For $N\lesssim 2$ we require a different approach and turn to the
consequences of assuming Gaussian stochastic behaviour.

\subsection{Gaussian correlations}

If $x$ measures distance along the annulus, $\phi (x)$ is periodic (modulo $2\pi$). The
 fluxon number density (or winding number density) is:
 \begin{equation}
 n(x) = \frac{1}{2\pi}\partial_x\phi(x),
 \end{equation}
 whereby the net fluxon number $n$ is:
 \begin{equation}
 n = \int_0^C\,dx\,n(x) = \frac{1}{2\pi}\Delta\phi,
 \end{equation}
 where $\Delta\phi$ is the change in $\phi$.

For winding number density $n\left(x\right)$ the ensemble average
of the net number of fluxons along an annulus of perimeter $C$, in
the absence of an external field  is ${\bar n} =\langle n\rangle =
0.$

 We do not need to adopt any
particular form for $n\left(x\right)$. In the light of our earlier
discussion we now assume that it is a {\it Gaussian} variable
until the transition is complete, whereby all correlation
functions are determined by the two-point correlation function
$\left\langle n\left(x\right)n\left(y\right)\right\rangle$. That
is, all we shall need for probabilities is:
\begin{eqnarray}
\langle n^2\rangle &=&\int_{0}^{C}dx\, dy \left\langle
n\left(x\right)n\left(y\right)\right\rangle .
 \label{g2}
 \nonumber
\end{eqnarray}
It follows that:
\begin{eqnarray}
\langle n^{2p}\rangle &=& \frac{(2p-1)!}{2^{p-1}(p-1)!}\langle
n^2\rangle^{p}.
 \label{eq:g2p}
\end{eqnarray}

\noindent If $f_m$ is the probability of finding net winding number
$m$ taking both positive and negative values), then:
\begin{equation}
 \langle n^{2p}\rangle =  \sum_{-\infty}^{\infty}m^{2p}\,f_m.
 \label{n2p0}
 \end{equation}
In order to invert equation (\ref{n2p0}), we construct the
generating function $Z(z)$:
\begin{eqnarray}
Z(z) &=& \sum_{p=0}^{\infty}\frac{(-z^2)^p}{(2p)!}\langle
n^{2p}\rangle = \exp (-z^2\langle n^2\rangle /2),
 \label{Z1}
  \nonumber
\end{eqnarray}
 from (\ref{eq:g2p}). On the other hand, from (\ref{n2p0}):
 \begin{eqnarray}
Z(z) &=& \sum_{p=0}^{\infty}\frac{(-z^2)^p}{(2p)!}
\sum_{-\infty}^{\infty}m^{2p}\,f_m
=\sum_{-\infty}^{\infty}f_m\,\cos mz.
 \label{Z2}
  \nonumber
\end{eqnarray}
 That is, the $f_m$ are the Fourier coefficients of the Gaussian.
In particular,
 \begin{eqnarray}
f_0 &=& \frac{1}{2\pi}\int^{\pi}_{-\pi}dz\,\exp (-z^2\langle
n^2\rangle /2)
 \label{fc0}
 \end{eqnarray}
 and
 \begin{eqnarray}
f_{\pm 1} &=& \frac{1}{2\pi}\int^{\pi}_{-\pi}dz\,\exp (-z^2\langle
n^2\rangle /2)\cos z
 \label{fcm}
 \nonumber
 \end{eqnarray}
\noindent respectively.

We observe that, for large $\langle n^2\rangle$, where we can take
the integration limits to infinity, the trapping probability falls
off as:
 \begin{equation}
 f_{\pm m} \approx \frac{1}{\sqrt{2\pi\langle n^2\rangle}}\exp (- m^2/2\langle
 n^2\rangle).
 \label{asymp}
  \nonumber
 \end{equation}

\noindent In practice, this assumption of a Gaussian stochastic
density can only be approximate, on two accounts. Less
significantly, from our earlier comments, it ignores the
non-linearities of the system. More importantly, it does not fully
accommodate the periodicity of the annulus, to which we shall return
later. Nonetheless, it will be apparent as to which results are
reliable.

In contrasting Gaussian correlations and Gaussian probabilities we
see that the definitions are dual to each other. In the limit of
unrestricted integration (large $N$) they are identical if we
identify:
 \begin{equation}
 <n^2>= \sigma^2(N)/4\pi^2 = N/12.
 \label{compo}
 \end{equation}
\noindent However, for small $N$ there will be differences, as we
shall see.

On inserting (\ref{compo}), we find that the likelihood ${\bar f}_1$
of seeing one fluxon or antifluxon is:
 \begin{eqnarray}
{\bar f}_1(N) &=& \frac{1}{\pi}\int^{\pi}_{-\pi}dz\,\exp (-z^2
N/24)\cos z,
 \label{fs1}
 \end{eqnarray}
 whereas, assuming Gaussian probability, from (\ref{fpm}) we find:
   \begin{equation}
 {\bar f}_1(N) =  erf (\frac{3\sqrt{3}}{\sqrt{2N}}) - erf
 (\frac{\sqrt{3}}{\sqrt{2N}}).
 \label{Fs1}
 \end{equation}

\noindent We know that (\ref{fs1}) and (\ref{Fs1}) agree for large
$N$ but, as can be seen from Figures 4, the qualitative and
quantitative agreement is striking over the whole range.

In Fig.4 we compare ${\bar f}_1$ as a function of $N$, as given by
Eq.(\ref{fs1}) and Eq.(\ref{Fs1}) (dashed and dotted lines)
respectively; the dots are the values of ${\bar f}_{1}$ according
to the independent sector model for some integer values of $N$
from Eq.(\ref{fmN}). Agreement is already good at $N=2$ and very
good at $N=4$.

\begin{figure}
 \hspace{2cm}\scalebox{0.8}
 {\includegraphics{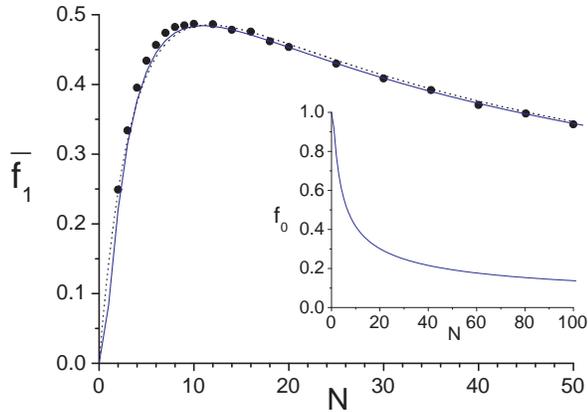}}
\caption{We display the probabilities ${\bar f}_1=f_{1}+f_{-1}$ as a
function of $N$, as given by Eq.(\ref{fs1}) and Eq.(\ref{Fs1})
(dashed and dotted lines) respectively; the dots are the values of
${\bar f}_{1}$ according to the independent sector model for some
integer values of $N$ from Eq.(\ref{fmN}). The maximum of ${\bar
f}_1$ occurs at $N=N_c\approx 10.5$, for which value ${\bar
f}_1\approx 49\%$. The inset shows the probability  ${\bar f}_0$ as
a function of $N$, as given by Eq.(\ref{fp0}), essentially
indistinguishable from that by Eq.(\ref{fc00}) at this scale.}
\label{Fig.4}
\end{figure}

 However, the plots of Fig.4 are somewhat deceptive  for small $N$.
  This should not worry us since, although the
 analytic expression (\ref{Fs1}):
   \begin{equation}
 {\bar f}_1(N) = O(\sqrt{N})\exp (-3/2N),\label{F1small}
  \end{equation}
 vanishes faster than any power, (\ref{Fs1})
breaks down there by definition. On the contrary, ${\bar f}_1(N)$ of
(\ref{fs1}) is {\it linear} in $N$ for small $N$,
 as we supposed in (\ref{flin}).

Finally, in the inset of Fig.4, we also display $f_0(N)$, the
probability of seeing no flux, for the case of Gaussian
probabilities given in (\ref{fp0}). Although we do not show it,  at
the scale of the plot, it is essentially indistinguishable from the
result of assuming Gaussian correlations as given in (\ref{fc0}),
 \begin{eqnarray}
f_0(N) &=& \sqrt{\frac{6}{\pi
N}}erf\bigg(\frac{\pi}{2}\sqrt{\frac{N}{6}}\bigg)\nonumber
 \label{fc00}
 \end{eqnarray}
 We can comfortably use either.
\subsection{Consequences}

In summary, the assumptions of Gaussian probabilities (as follows
from the KZ picture) and Gaussian correlations are complementary,
with the latter providing a (linear) interpolation of the other
for $N\lesssim 2$ where the former breaks down.

 The first observation is that, in
both cases, the maximum probability ${\bar f}_1 = {\bar f}_c$ of
seeing one fluxon {\it or} antifluxon is fractionally less than
50\% (48.6\%), occurring at $N = N_c\approx 10.5$. The assumption
made in our JTJ papers is that ${\bar f}_1$ (which we have called
$f_1$ in our papers) scales linearly with $N\propto C/\xi$, which
we see is valid at best only until ${\bar f}_1 \approx 0.3$.  We
have already just about achieved this in the existing experiments
but have not yet been able to quench fast enough to provide a
direct test of the model predictions of Fig.4. There is, however,
a problem that stops us embracing the Gaussian correlation
approach wholeheartedly, as we have presented it here. As we have
noted, the assumption of Gaussian winding number density can only
be approximate, since it does not take periodicity (mod $2\pi$)
into account. We have seen that this does not matter in the
calculation of $f_0$ and ${\bar f}_1$.

However, for small annuli there is a  problem with (\ref{fs1}) in
that Fourier components $f_{\pm 2}(N)$
 are slightly {\it negative} ($|f_{\pm 2}|< 0.01$) for very {\it small} $N$ (with a similar problem for the very
much smaller ${\bar f}_4$).  On the other hand, by construction
the ${\bar f}_m(N)$ of (\ref{Fs1}) are automatically positive as
they must be.

In a qualitative sense it is of little consequence since, throughout
the linear regime, the probability ${\bar f}_> = 1 - f_0 - {\bar
f}_1$, of seeing more than one fluxon is very small but, as a matter
of principle, we should impose periodicity to render the
probabilities positive.
 We do not have a reliable model in which we can do this but we
 can get some idea by supposing that the phase $\phi$ moved in a
 double-well potential rather than the periodic $\cos\phi$
 potential of the sine-Gordon fluxon. In that case, assuming Gaussian correlations, the density of
 defects is proportional \cite{halperin} to $(f''(0)/f(0))^{1/2}$, where $f(x) =
 \langle\phi(x)\phi(0)\rangle$. It is now
 straightforward to impose periodicity, whereupon we find
 behaviour similar to that of (\ref{F1small}) for very small $N$, in that it vanishes
 faster than any power \cite{Bob}. If that were to be equally applicable here it is difficult to determine
 when such behaviour might occur, since there is no sign of such a collapse in Fig.1,
 but if this analogy is correct, it will repair the minor problem
 of small negative probabilities
 while leaving the similarity between the two approaches at a
 quantitative level. In particular, as long as we are not looking at the small-N
 behaviour in too much detail, as we shall not hereafter, it becomes sensible to use the simpler
 Gaussian probabilities over the whole range.

\section{Fluxon Production in an external field $B\neq 0$}

Let us now  apply a perpendicular uniform magnetic field $B$ to
the AJTJ. This breaks the $\phi\rightarrow -\phi$ symmetry that is
equally, the reflection symmetry of the system in the plane of the
barrier. Once superconducting, the AJTJ expels the magnetic field,
but we assume that a small fraction $\epsilon$ of the applied
field 'leaks' in the radial direction through the barrier, forming
fluxons. The effect of this field is to produce an non-zero
average winding number $\langle n\rangle = {\bar n}(B)$.

The result is a shift in phase gradient along the annulus
(coordinate $x$) of the form:
 \begin{equation}
 \partial_x\phi \rightarrow \partial_x\phi + \frac{2e}{\hbar
 c}A_x,
 \end{equation}
where $A_x= A_x^+ - A_x^-$ is the jump in the vector potential
across the oxide layer.

If $C = 2\pi R$ is the circumference of the ring, radius R, then
 the change in $\phi$ due to $B$ is:
 \begin{equation}
 \Delta\phi = \epsilon\frac{2e}{\hbar c}\oint {\bf dl}.{\bf A} = \frac{\epsilon}{2\pi}\frac{e}{\hbar c} C^2B.
 \end{equation}
 The change in fluxon number is:
 \begin{equation}
 {\bar n}(B) = \frac{1}{2\pi}\Delta\phi =
 \frac{\epsilon}{4\pi^2\hbar c}e C^2B.\label{Nbar}
  \end{equation}

\noindent (For future purposes, let us call $B_1$ the field value
for which $\Delta\phi = {2\pi}$, that is ${\bar n}(B_1) =1$.)When
${\bar n}$ is small, $\epsilon$ adjusts in any individual
experiment so as to make ${\bar n}$ integer. As a first
approximation we take the fraction $\epsilon$ to be independent of
$B$.

\noindent In the presence of external fields we are not primarily
interested in the small $N$ linear regime and it is sufficient to
work with Gaussian probabilities. The natural extension of $G(\Phi)$
of (\ref{Ggauss}) for an AJTJ in a perpendicular magnetic field $B$
is that the phase distribution will still be normal with variance
$\sigma^2(N,{\bar n}(B))\propto N$, where we retain the definition
of $N\propto C/{\bar\xi}$ of the previous section, but with non-zero
average $\bar\Phi(B) = 2\pi{\bar n}(B)$:

\begin{equation}
G_{N,\bar n}(\Phi) = \frac{1}{\sqrt{2 \pi \sigma^2(N,{\bar n})} }
\exp-\frac{[(\Phi-\bar\Phi(B)]^2}{2 \sigma^2(N,{\bar n})}.
\label{Ggauss2}
 \nonumber
\end{equation}

\noindent The trapping probabilities in the presence of an external
magnetic field will then be:
\begin{equation}
f_{m}(N,\bar n) = \int_{-\pi + 2m\pi}^{\pi +
2m\pi}d\Phi\,G_{N,\bar n}(\Phi); \label{fpq}.
\end{equation}

\noindent To a first approximation we assume that $\sigma^2(N,\bar
n)$ is independent of $\bar n(B)$, as would follow from assuming
Gaussian correlations, i.e.
 for integer $\bar n$, $f_{{\bar n} - m}(N,{\bar n}) = f_{{\bar n} + m}({N,\bar n}) = f_{\pm
m}(N,0)$.  This allows us to repeat the identification
$\sigma^2(N)/4\pi^2 = N/12$.

As before, we are primarily interested in the probability of seeing
a single fluxon or antifluxon, but now for fixed $\tau_Q$ (or $N$),
as a function of $B$ (or ${\bar n}$). Now, as far as $\bar n \neq
0$, then the symmetry is broken and $f_{+1}(N,{\bar n})\neq
f_{-1}(N,{\bar n})$. More precisely, in Fig. 5 we show ${\bar
f}_1(N, {\bar n})$ as a function of ${\bar n}$ for fixed $N$ for
several $N$, as derived from Gaussian distributions Eq.(\ref{fpq}).
\begin{figure}[htp]
\begin{center}
\epsfysize=7cm \epsfbox{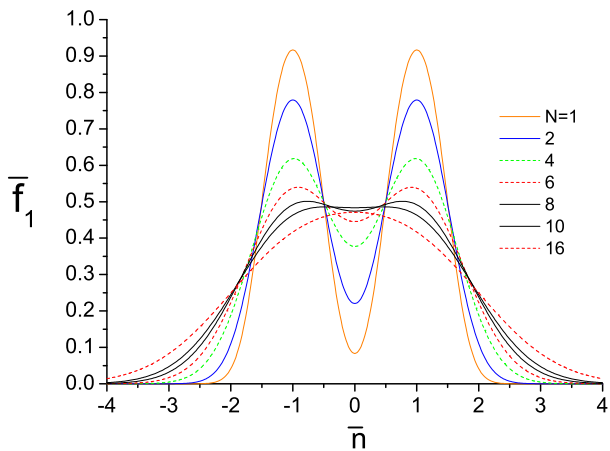}
\end{center}
\caption{The values of ${\bar f}_1(N,\bar n)$ as a function of
$\bar n$ for fixed $N$ for several values of $N$; $N = 1, 2, 4, 6,
8, 10, 16$, according to Eq. (\ref{fpq}).} \label{Fig.5}
\end{figure}

The main characteristics of Fig.5 are that:

\begin{enumerate}
\item for $N <N_c\approx 10.5$ there is a double peak,
corresponding to the ensemble average production of a single
fluxon by the applied external magnetic field. This is understood
as follows. Essentially, when, for $B=B_1$ we have ${\bar n}=1$,
the zero-field no-trapping frequency $f_0(N,0)$ becomes the single
fluxon trapping $f_{+1}(N,1)=f_0(N,0)$, giving the righthand peak
for positive ${\bar n}\approx 1$; reversing the field,
$f_{-1}(N,-1)=f_0(N,0)$ and we get the peak for negative ${\bar n}
\approx -1$.The minimum value of ${\bar f}_1$ between the peaks is
${\bar f}_1(N)$, as given by the KZ scenario.
 \item
  as $N$ increases to
$N_c$ the height ${\bar f}_1(N,0)$drops. The variance also
increases, the $\Phi$ distribution gets broader and the two peaks
in $\bar n$ of ${\bar f}_1(N,{\bar n})$ merge at $N= N_c$ at the
value ${\bar f}_c\approx 0.5$;
 \item
 as $N$ increases beyond $N_c$, there is only a single peak
centered on $\bar n = 0$.  We now see that the reason why the
probability of seeing a fluxon {\it decreases} as $|B|$ increases
is a consequence of an increased ability to create more than one
fluxon.
 \end{enumerate}
\noindent The curves in Fig.5 for $N=1$ and $N=16$ bear a strong
resemblance to the experimental data shown in Fig.2 and Fig.3,
respectively. They comply with the most simple qualitative test of
our analysis, that the double peaks in Fig.1 occur at a {\it
higher} frequency than ${\bar f}_c\approx 50\%$, and the single
peak in Fig.2 at a {\it lower} frequency than $50\%$, as
predicted. Further, we also found that the defect winding number
flips when we move from the left to the right peak, as foreseen by
both models. This discrimination between a trapped fluxon or
antifluxon can be achieved by measuring the transverse magnetic
field dependence of the junction critical current $I_c(B)$ (the
details of this new effect and its theoretical interpretation will
be reported elsewhere \cite{Monaco6}).

\noindent More specifically, we note that the $2.0mm$ long sample,
being $4$ times longer than the $0.5mm$ long sample, according to
(\ref{Nbar}), should be 16 times more sensitive to the externally
applied magnetic field $B$, if $\epsilon (B)$ is independent of $B$
and identical for both samples. There is, indeed, a strong
difference in sensitivity, but only by a factor of $7-8$, showing
that these assumptions are approximate. What is more difficult to
understand quantitatively is the 16-fold increase in $N$ for a
four-fold increase in perimeter. This requires the efficiency factor
$a$ relating $N$ to $C$ to vary by a factor of $4$ between the
samples or, more fundamentally, that $N$ is not {\it linear} in $C$.
Of itself, the latter does not change the scaling behaviour of
(\ref{flin}), but the scaling exponent $\sigma$. In
\cite{Monaco3,Monaco4} we showed that the observed value for
$\sigma$ was not that for idealised JTJs \cite{KMR}. We explained
this as a consequence of fabrication methods, but this reopens the
issue.

 To go
further, and match the data profiles better, we need specific
properties of JTJs, beyond the generics of the KZ picture (or
Gaussian correlations). In particular, the assumption of
$\sigma^2(N,{\bar n}(B))$ being independent of ${\bar n}(B)$ is
oversimple. More realistically \cite{Martucciello},
$\sigma^2(N,{\bar n}(B))= \sigma^2(N,0) (1 + k^2 B^2)$, where
$k^2\propto (\lambda_J/ C)^2$, i.e. it is inversely proportional
to the ring area and to the Josephson current density $J_c$, since
$\lambda_J^2\propto 1/J_c$. Both are consistent with the
experimental data, such as the flattening of ${\bar f}_1$ for
small $B$ field values and the fact that the peak amplitudes
$f_{\pm n}(\pm B_n)$ strongly decrease with $n$, as we shall see
elsewhere.

\section{Conclusions}

We have developed two complimentary theoretical approaches to
understand the experimentally observed spontaneous production of
fluxons on quenching annular JTJs in the presence of an externally
applied transverse symmetry breaking magnetic field $B$. They
either assume Gaussian probabilities (as motivated by KZ causal
horizons) or Gaussian correlation functions (as motivated by
models for transitions based on the rapid growth of
instabilities). Both of these approaches, which are,
approximately, identical, lead to the same scaling behaviour of
(\ref{xibar}), from which (\ref{flin}) follows in the appropriate
regime.

 The theory is able to nicely
reproduce the double peak behavior of the likelihood $f_1$ to
produce a single defect(fluxon) shown in Fig.2 for a sample having a
circumference equal to $0.5 mm$ and the single peak in Fig.3 for a
sample of circumference $2.0 mm$. When we began experiments on
fluxon production in an external symmetry-breaking field, we
anticipated the behaviour shown in Fig.2, and not that of Fig.3,
which was initially incomprehensible. We now understand it, as a
consequence of the ease of producing more than one fluxon in larger
annuli. Specifically, our models do provide a good first
approximation at a better than qualitative level and the
experimental success of the Gaussian picture in describing the
production of fluxons is of a piece with the scaling behaviour so
robustly demonstrated in Fig.1.

 \section*{Acknowledgements}

We thank Arttu Rajantie for helpful discussions. RM acknowledges the
support of ESF under the COSLAB project and CNR under the Short-Term
Mobility program.

\end{document}